\documentclass[pra,letterpaper,twocolumn,showpacs,superscriptaddress,floatfix]{revtex4}

\usepackage{graphicx,psfrag,amsmath,amssymb,amsfonts,bbm,latexsym,color,dcolumn,bm}

\begin{document}

\title{A proposal for a quantitative indicator of original research output}

\author{Roberto Onofrio}

\affiliation{\mbox{Dipartimento di Fisica e Astronomia ``Galileo Galilei'', Universit\`a  di Padova, 
Via Marzolo 8, Padova 35131, Italy}}

\affiliation{\mbox{Department of Physics and Astronomy, Dartmouth College, 6127 Wilder Laboratory, 
Hanover, NH 03755, USA}}

\begin{abstract}
The use of quantitative indicators of scientific productivity seems now quite widespread
for assessing researchers and research institutions. There is a general perception, however, that these
indicators are not necessarily representative of the originality of the research carried out, being
primarily indicative of a more or less prolific scientific activity and of the size of the
targeted scientific subcommunity. We first discuss some of the drawbacks of the broadly adopted
$h$-index and of the fact that it represents, in an average sense, an indicator derivable from the
total number of citations. Then we propose an indicator which, although not immune from biases, seems
more in line with the general expectations for quantifying what is typically considered original work.
Qualitative arguments on how different indicators may shape the future of science are finally discussed.
\end{abstract}


\maketitle

\section{Introduction}
The need to assess scientific productivity has always been present in the history of science, and
it plays an ever growing role considering the budgetary constraints that have emerged since almost
three decades of post-cold-war science.
A single-number indicator of scientific impact, named $h$-index, was proposed more than a
decade ago, and since then has been progressively considered as relevant especially for
hiring, tenure, and promotion processes in many academic systems \cite{Hirsch}.
In spite of the caveats explicitly discussed in detail by his originator, the $h$-index 
is currently considered the most crucial indicator for science output, and is affecting significantly
the evolution of entire research fields.
There is an extensive literature on the shortcomings of the $h$-index, and various modifications
have been proposed to address these issues (see for instance \cite{Hirschbar} and references [4-21] therein). 
Here we show that the $h$-index is, in an average sense, a secondary indicator, merely representing a number
related to the total number of citations.
Moreover, we show that apparently paradoxical features emerge by considering deviations
from the average $h$-index. We then introduce an index more related than the $h$-index to the
degree of originality of scientific research and the perception that we have of outstanding scientists,
providing two examples from last century physics. While we warn for the general danger of adopting
single-number indicators as the main parameter for decision-making in science policy, we finally argue
that adoption of the proposed $\Omega$-index may result in a quite different historical evolution of
various research fields with respect to the one determined by the $h$-index.

\section{The $h$-index as a secondary indicator}

The $h$-index, the integer number for which $h$ papers have collected at least $h$ citations, is considered 
a simple, effective way to quantify with a single number the impact of an individual or an institution.
The $h$-index is contracting the full information of a histogram in which the number of citations of each paper
is plotted versus the progressive number of paper enumerated in decreasing order of citations, the ``citation curve''
$y=N_{\mathrm{cit}}(i)$, and is obtained by intersecting this histogram with the line $z=i$ \cite{Hirsch}.

\begin{figure}[t]
\includegraphics[width=1.00\columnwidth, clip=true]{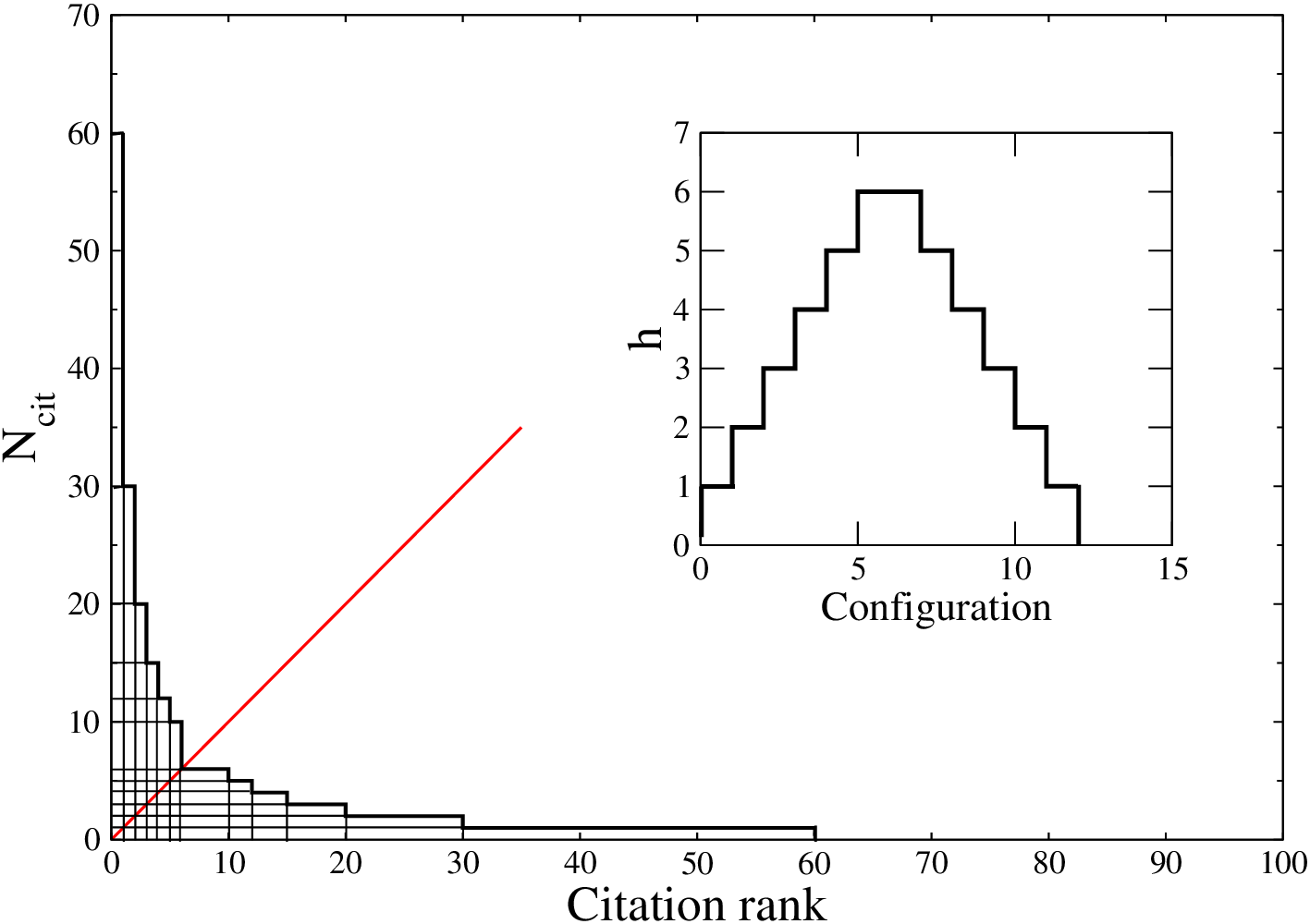}
\caption{Configurations sharing the same total number of citations in the 'rectangular' approximation
discussed in the text. The $h$-index is obtained by intersecting each configuration with the red straight line.
The resulting $h$-index for each configuration appears in the inset, showing that both configurations
corresponding to high and low impact papers indiscriminately result in a small $h$-index.
The example refers to an hypothetical author with a total of 60 citations spread over 60 papers, with
the following configurations corresponding to only integer numbers for both citations and papers:
1 paper with 60 citations (1), 2 papers with 30 citations (2), 3 papers with 20 citations (3), 4 papers with
15 citations (4), 5 papers with 12 citations (5), and 6 papers with 10 citations (6), and the mirrored
configurations (7-12) obtained exchanging number of papers and citations. Summing the $h$-indexes over all
the configurations we obtain $h_{\mathrm{av}}$=3.5. The same value is obtained if the number of papers is
$N_p={\mathrm{int}}(\sqrt{N_\mathrm{cit}})$ as this means to average over half of the configurations on the inset.
In the intermediate situation of ${\mathrm{int}}(\sqrt{N_{\mathrm{cit}}})< N_p < N_{\mathrm{cit}}$ the value of $h_{\mathrm{av}}$ is
larger than 3.5.}   
\label{Fig1}
\end{figure}

A number of criticisms has been raised against the capability of the $h$-index to adequately represent
scientific accomplishments, some of them already discussed at its very inception. While the complete
information on the scientific output is available through the citation curve \cite{Peterson,Petersen},
the $h$-index is a coarse-grained indicator which does not provide information neither on the papers with
high number of citations, nor on the tail consisting of papers with low number of citations: By construction, all
the relevant information of the $h$-index is played at the intersection point of the $y$ and $z$ curves.
Moreover, biases can be introduced artificially by self-citations, as researchers willing to
boost their $h$-index will intentionally cite their papers at the cross-over region.
By just self-citing all former papers in the same field in a new contribution, a steady increase
of the $h$-index is guaranteed \cite{Schreiber}. Therefore, the $h$-index penalizes researchers switching
to different fields with respect to the ones continually pursuing the same problem and presumably
citing their former contributions in the same field. 
This issue could be easily corrected in the analysis of most databases for which self-citations 
are available, although this has not been systematically implemented so far. 
There are other issues, however, that cannot be easily addressed with the $h$-index. The size of the relevant
scientific community is indeed very crucial, so problems interesting few researchers, no matter how potentially relevant 
and challenging they might be, are downplayed with respect to mainstream research involving larger numbers
of researchers and papers. Since, typically, the size of scientific communities increases in time,
researchers who built up a productive career several decades ago are somewhat penalized with respect
to the newer generations. Of course, an indicator which depends on the size of the community and on
time can hardly be suitable to compare different fields or researchers with different seniority and interests,
and alternative measures have been discussed to overcome this strong limitation \cite{Pepe}. 

\begin{figure}[t]
\includegraphics[width=0.98\columnwidth, clip=true]{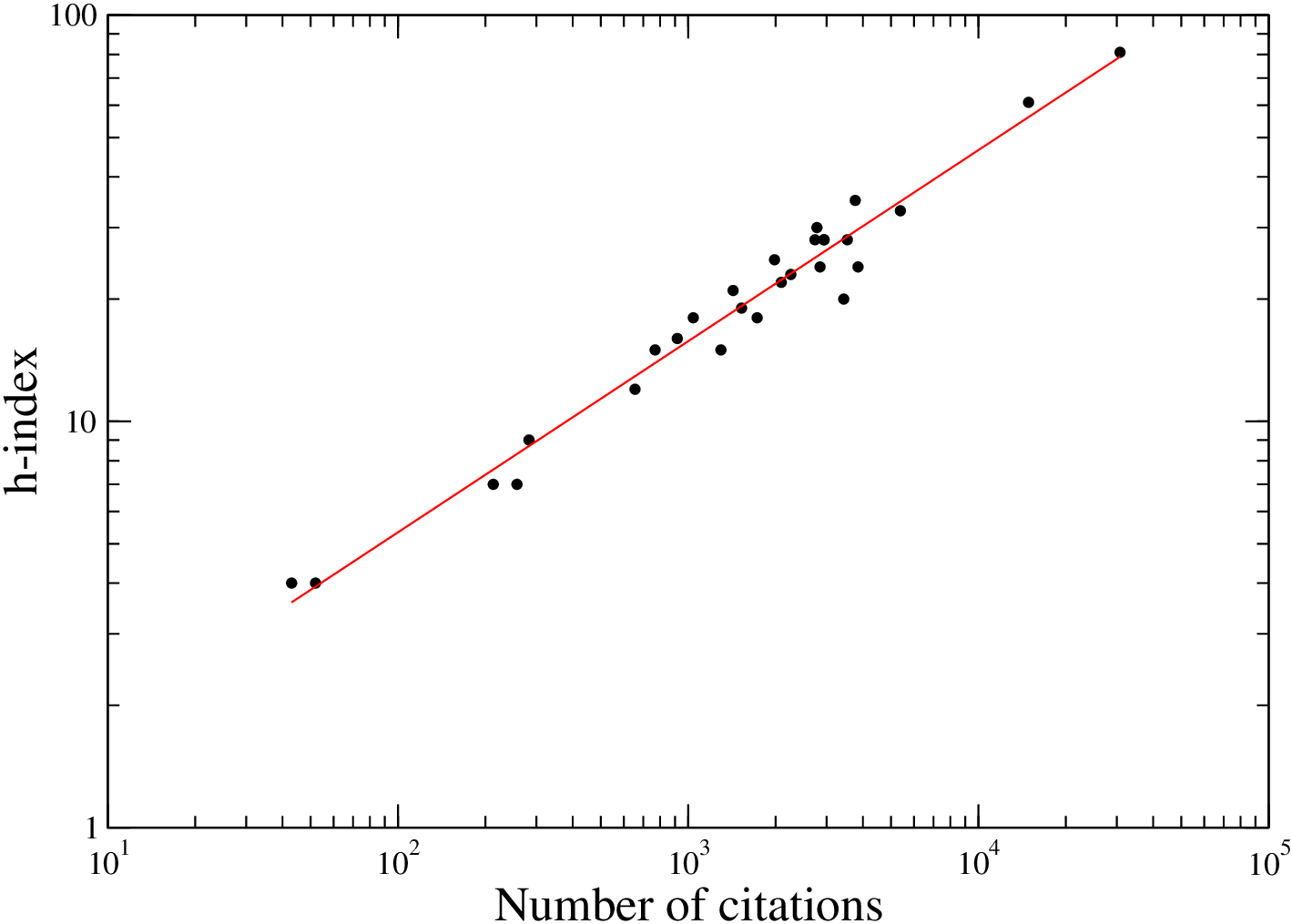}
\caption{Plot of the $h$-index versus the number of citations for the authors and collaborators (black circles),
limited to the ones having coauthored at least four papers, and best fit with a power-law function (red line),
$h= \alpha N_{\mathrm{cit}}^\beta$, with $\alpha=0.62 \pm 0.08$ and $\beta= 0.47 \pm 0.02$,
consistent within two standard deviations with the prediction of Eq. (\ref{Eq1}).
All data reported here and in the following figures and tables have been extracted from the 
Web of Science database in the timeframe between 10/17/2017 and 10/23/2017, selecting the
Science Citation Index Expanded setting alone, thereby excluding conference proceedings.}
\label{Fig2}
\end{figure}

We now show that there is a strong correlation between the total number of citations $N_{\mathrm{cit}}$
and the $h$-index.
An empirical relation was already identified in \cite{Hirsch}, as $N_{\mathrm{cit}}=a~ h^2$, with $a$ a parameter in the
3-5 range. We would like to derive from first principles a relationship between $N_{\mathrm{cit}}$ and the
expected $h$-index. In order to reach this goal, it is worth to stress that, for a given total
number of citations $N_{\mathrm{cit}}$, various $h$-indexes may be achieved.
In one extreme case, it is possible to have a histogram in which one paper gets all $N_{\mathrm{cit}}$ and
all others get zero citations. Next to this, it is possible that two papers are instead cited, and all
the other are not, for instance one with $N_{\mathrm{cit}}-1$ citations, the other with one citation. 

In general, the number of possible distinct configurations given a number of citations coincides with the evaluation 
of the partitions of $N_{\mathrm{cit}}$. This number increases quickly with $N_{\mathrm{cit}}$ -- exponentially according
to the asymptotic estimate of Hardy, Ramanujan and Uspensky -- for instance for 
$N_{\mathrm{cit}}=100$ there are 190,569,292 partitions, while for $N_{\mathrm{cit}}=1000$ there are
$\simeq 2.40615 \times 10^{31}$ partitions. 
Each partition has an associated $h$-index, and one could obtain an average $h$-index by averaging the $h$-indexes
over all the partitions, for instance assuming that each has the same probability to occur {\it a priori}.
This becomes intractable for the typical values of available $N_{\mathrm{cit}}$, and therefore we adopt an approximation
consisting in limiting the number of configuration to ``rectangular'' patterns in which $k$ papers share the same number
of $N_{\mathrm{cit}}/k$ citations, see Fig.~\ref{Fig1} for an example. We will consider only a situation in which
a configuration has only one paper cited $N_{\mathrm{cit}}$ times (corresponding to $h=1$), another in which
two papers get $N_{\mathrm{cit}}/2$ (corresponding to $h=2$), etc. This goes all the way to the other extreme
case in which one has $N_{\mathrm{cit}}/N_p$ citations in all the $N_p$ papers. In the specific case in which one has
$N_{\mathrm{cit}}=N_p$, and if all configurations are equally probable, we have to sum them with equal
weight to achieve the average $h$-index, named $\langle h \rangle$ thereafter. Considering the symmetry
of the configurations around the point with coordinates
$({\mathrm{int}}(\sqrt{N_{\mathrm{cit}}}),{\mathrm{int}}(\sqrt{N_{\mathrm{cit}}}))$ with
${\mathrm{int}}$ indicating the integer part of $\sqrt{N_{\mathrm{cit}}}$, we have

\begin{equation}
\langle h \rangle= \frac{1+2+... {\mathrm{int}}(\sqrt{N_{\mathrm{cit}}})}{{\mathrm{int}}(\sqrt{N_{\mathrm{cit}}})}= 
\frac{1+{\mathrm{int}}(\sqrt{N_{\mathrm{cit}}})}{2} \approx \frac{1}{2}\sqrt{N_{\mathrm{cit}}}.
\label{Eq1}      
\end{equation}
with the last expression holding for $N_{\mathrm{cit}} \gg 1$. Therefore, the average $h$-index,
$\langle h \rangle$, at least under the two hypotheses made, {\it i.e.} the ``rectangular'' one
and the fact that $N_{\mathrm{cit}}=N_p$, is nothing but a reparametrization of the total number of
citations, just differing from it in being sensitive to the details of the citation curve at the crossover point.
The empirical factor $a$ introduced in \cite{Hirsch} is recovered, at its central value of $4$, in our rectangular
approximation. Figure ~\ref{Fig2} show an example suggesting that $\langle h \rangle$ evaluated as in Eq. (\ref{Eq1}) is
a good interpolation of real data, evaluated from the personal experience for a variety of researchers of different
seniority, all sharing average achievements, and for a broad range of total citations, ranging from postgraduate
students who left academic research immediately after their studies to a Nobel laureate.

Interesting considerations may be made by looking at {\it deviations} of the actual data from $\langle h \rangle$,
which may be related to the variability of the $a$ parameter noticed in \cite{Hirsch}. For the same total number of
citations, researchers with an actual $h$-index significantly smaller than the corresponding $\langle h \rangle$
have few papers with high number of citations, and many papers with marginal impact, with the extreme case of
all citations available in a single paper. At the opposite end of the spectrum lies instead a class of researchers
with several medium-impact papers, but no papers with large numbers of citations, with $\sqrt{N_{\mathrm{cit}}}$ papers
each having $\sqrt{N_{\mathrm{cit}}}$, and all others with zero citations. The typical presence of more citations than
papers, especially for more senior researchers, breaks the symmetry of configurations on which Eq. (\ref{Eq1}) is
derived, weighting more the lower configurations with large number of citations. This corresponds to the case
of $a$ closer to smaller values than the middle value of 4 in the empirical remark reported in \cite{Hirsch}. 

The case of researchers having $h \ll \langle h \rangle$ represents individuals pursuing high-risk/high-reward work,
while the opposite case is more representative of researchers who, in Kuhn's approach, are more inclined to work on
{\sl normal} science. In the perception of the scientific community and the outside world, a successful scientist
is typically more associated to the former case \cite{Redner,Bagrow}.
By looking at history of physics, it turns out that most of the founding fathers of quantum
mechanics have in fact $h$-indexes well {\em below} the corresponding average one.
In more recent times, there is the striking case of Nobel laureate Peter Higgs who, according to the
Web of Science, has a $h=11$ with 22 papers totaling 5642 citations as of 10/17/2017.
Meanwhile there are various researchers who have $h$-index
in the 80-100 range and who, although well-known in their community, are not necessarily perceived
as belonging to the same category of creative thinkers as Higgs. They are rather considered as belonging to
a category of prolific scientists, may be also due to the large size of their research groups or
the sheer size of the community to which they belong. A question arises naturally: As the $h$-index seems
to be anticorrelated with breakthrough results and original work, is it possible to find another indicator
overcoming this limitation? While a rather sophisticated indicator has been already proposed
\cite{Soler}, in the next section we will try to give a simpler answer to this question.

\section{An indicator suited for scientific originality}

The assessment of creative work is not univocal, and creativity is very often rather elusive and
intangible, see for instance artist communities in which the consensus on creativity is
often controversial, with the debate sustaining an entire sector, those of the critics, for centuries.
Late assessments, while crucial in history and sociology of science, are manifestly of marginal interest
for academic comparison of candidates. However, if one insists in aiming to quantify originality with
a single number, a simple indicator may be constructed as follows.
We introduce the originality index $\Omega$ as the ratio

\begin{equation}
\Omega=\frac{N_{\mathrm{cit}}-N_{\mathrm{ref}}}{N_{\mathrm{ref}}},
\label{Eq2}
\end{equation}
where $N_{\mathrm{cit}}$, as before, is the total number of citations of all the papers of a given individual 
or institution, and $N_{\mathrm{ref}}$ is the total number of references cited in the same papers. 
Qualitatively, one may imagine that any scientific result emerges from previous knowledge, 
here quantified by the amount of references cited in each contribution. 
The success, in terms of originality, of the new result will be measured by the ability to create
new scholarship, measured in turn by the citations obtained with respect to the existing scholarship. 
In the example of Peter Higgs, the apparently modest $h$-index of $11$ is
contrasted by an $\Omega$=21.04, thanks to a total of only 256 references in his papers.

There are various advantages of this indicator. As the $h$-index, it is simple to evaluate.
The total number of citations is already available in various databases, and it needs to be
complemented by a count on the total number of references. The latter does not need to be updated
until a new published paper is added to the database. If the average number of citations per
paper $n_{\mathrm{cit}}$ and the average number of references per paper $n_{\mathrm{ref}}$ are available,
the originality index is simply expressed in terms of these intensive parameters as
$\Omega=n_{\mathrm{cit}}/n_{\mathrm{ref}}-1$. Self-citations are automatically canceled out as, by definition, 
they contribute to both quantities in the numerator of Eq. (\ref{Eq2}). 
The production of papers of secondary relevance will be inhibited if one seeks to maximize $\Omega$, as
they will most likely end up in adding many more references than future citations.
Likewise, the proliferation of secondary scholarship with {\it a priori} little originality ({\it e.g.}
review articles), will also affect negatively the originality index at least as long as the intrinsically
large number of references will not be offset by a proportionate number of citations.
\begin{table}[t]
\begin{center}
{\scriptsize
  \begin{tabular}{|l|c|c|c|c|c|l|}
\hline\hline
Author             & $N_{\mathrm{p}}$  & $N_{\mathrm{cit}}$ & $N_{\mathrm{ref}}$ & $h$ & $\langle h \rangle$ & $\Omega$ \\
\hline
Bohr,N.           & 79  & 6835   & 730         & 29  & 41.3  &  8.36        \\
Born,M.           & 192 & 14471  & 1780        & 42  & 60.1  &  7.13        \\
Bragg,W.L.        & 98  & 4105   & 643         & 37  & 32.0  &  5.38        \\
Brillouin,L.      & 74  & 2298   & 591         & 20  & 24.0  &  2.89        \\
Compton,A.H.      & 110 & 2157   & 1163        & 23  & 23.2  &  0.86        \\
Curie,M.          & 25  & 145    & 502         & 5   &  6.0  & -0.71        \\
de Broglie,L.     & 120 & 1229   & 374         & 14  & 17.5  &  2.29        \\  
de Donder,T.      & 29  & 33     & 86          & 3   &  2.9  & -0.62        \\
Debye,P.          & 153 & 18800  & 1198        & 52  & 68.6  & 14.69        \\
Dirac,P.A.M.      & 98  & 21510  & 320         & 46  & 73.3  & 66.22        \\       
Ehrenfest,P.      & 60  & 1355   & 610         & 18  & 18.4  &  1.22        \\
Einstein,A.       & 129 & 30171  & 186         & 53  & 86.8  & 161.21       \\
Guye,C.E.         & 22  & 10     & 17          & 2   & 1.6   & -0.41        \\ 
Heisenberg,W.     & 113 & 9790   & 1346        & 41  & 49.5  & 6.27         \\
Henriot,E.        & 12  & 73     & 68          & 4   & 4.3   &  0.07        \\
Herzen,E.         & 1   & 1      &  1          & 1   & 0.5   &  0.00        \\
Knudsen,M.        & 23  & 1614   & 104         & 14  & 20.1  & 14.52        \\
Kramers,H.A.      & 61  & 12958  & 653         & 30  & 56.9  & 18.84        \\
Langevin,P.       & 29  & 2393   & 184         & 11  & 24.5  & 12.01        \\
Langmuir,I.       & 154 & 30138  & 1994        & 58  & 86.8  & 14.11        \\
Lorentz,H.A.      & 46  & 434    & 133         & 8   & 10.4  & 2.26         \\
Pauli,W.          & 41  & 4233   & 443         & 27  & 32.5  & 8.56         \\
Piccard,A.        & 25  & 105    & 71          & 6   & 5.1   & 0.48         \\
Planck,M.         & 90  & 1388   & 454         & 13  & 18.6  & 2.06         \\
Richardson,O.W.   & 146 & 693    & 1131        & 12  & 13.2  & -0.39        \\
Schroedinger,E.   & 86  & 8844   & 853         & 28  & 47.0  & 9.37         \\
Verschaffelt,J.E. & 33  & 25     & 220         & 3   & 2.5   &-0.89         \\
Wilson,C.T.R.     & 34  & 1296   & 119         & 16  & 18    & 9.89         \\
\hline
\end{tabular}
}
\end{center}
\caption{\label{table1}
Rankings of the participants to the fifth Solvay Conference on Physics in 1927 according to data
available on Web of Science. Names of the participants in alphabetic order, number of published
papers through their whole carreer $N_p$, total number of citations $N_{\mathrm{cit}}$, total number
of cited papers $N_{\mathrm{ref}}$, $h$-index, average $h$-index, and $\Omega$-index are reported.}
\end{table}

The $\Omega$-index is also robust with respect to another possible phenomenon, authorship 
fusion, which instead may be favorable, barring obvious ethical principles, having in mind maximization
of the $h$-index. Suppose that there are two disjoint groups of authors $A$ and $B$ for two papers, their
$h$-index can be increased if the authors agree to share authorship, as this will allow to collect
more citations on the common paper. This gregarious author clustering is, instead, mitigated if the
$\Omega$ index is to be maximized, because the presence of a distinct set of references
$N_{\mathrm{ref}}^{(A)}$ and $N_{\mathrm{ref}}^{(B)}$ will make the fusion less advantageous, unless of course the
two sets will coincide, in which case it may indicate that actually the two groups of authors
have at least genuinely common research interests. The fusion phenomenon is particularly striking
for researchers who suddenly switch from a few-collaborator production mode to become members of
a large, inclusive collaboration. It is easy to observe that researchers who opted to join a large
research collaboration, as it happens for instance in astronomy and high-energy physics, experience
a sudden increase in the number of citations and the related $h$-index, with respect to people persisting
with a small-science mode of production.

Maximization of the $\Omega$-index is primarily obtained by switching to new problems, instead of staying
on the same territory for a long time which will maximize the $h$-index as discussed above.
Perhaps one of the most iconic example of this optimization is provided by Enrico Fermi, who
has $h$=41 with 82 papers, $N_{\mathrm{cit}}$=10,904, and $N_{\mathrm{ref}}$=515, corresponding to $\Omega$=20.17.
Most notably, Fermi has 17 papers containing no references, among these the first report on ``Radioactivity
induced by neutron bombardment'' (cited 75 times), and the paper
introducing a model for weak interactions with 10 references and 648 citations (with no other
follow-up paper written on the same subject). While the $h$-index of Fermi does not seem particularly
impressive today, the large $\Omega$-index speaks a lot about his agility to give original and impactful
contributions in a broad range of subfields of physics.
\begin{figure}[t]
\includegraphics[width=1.00\columnwidth, clip=true]{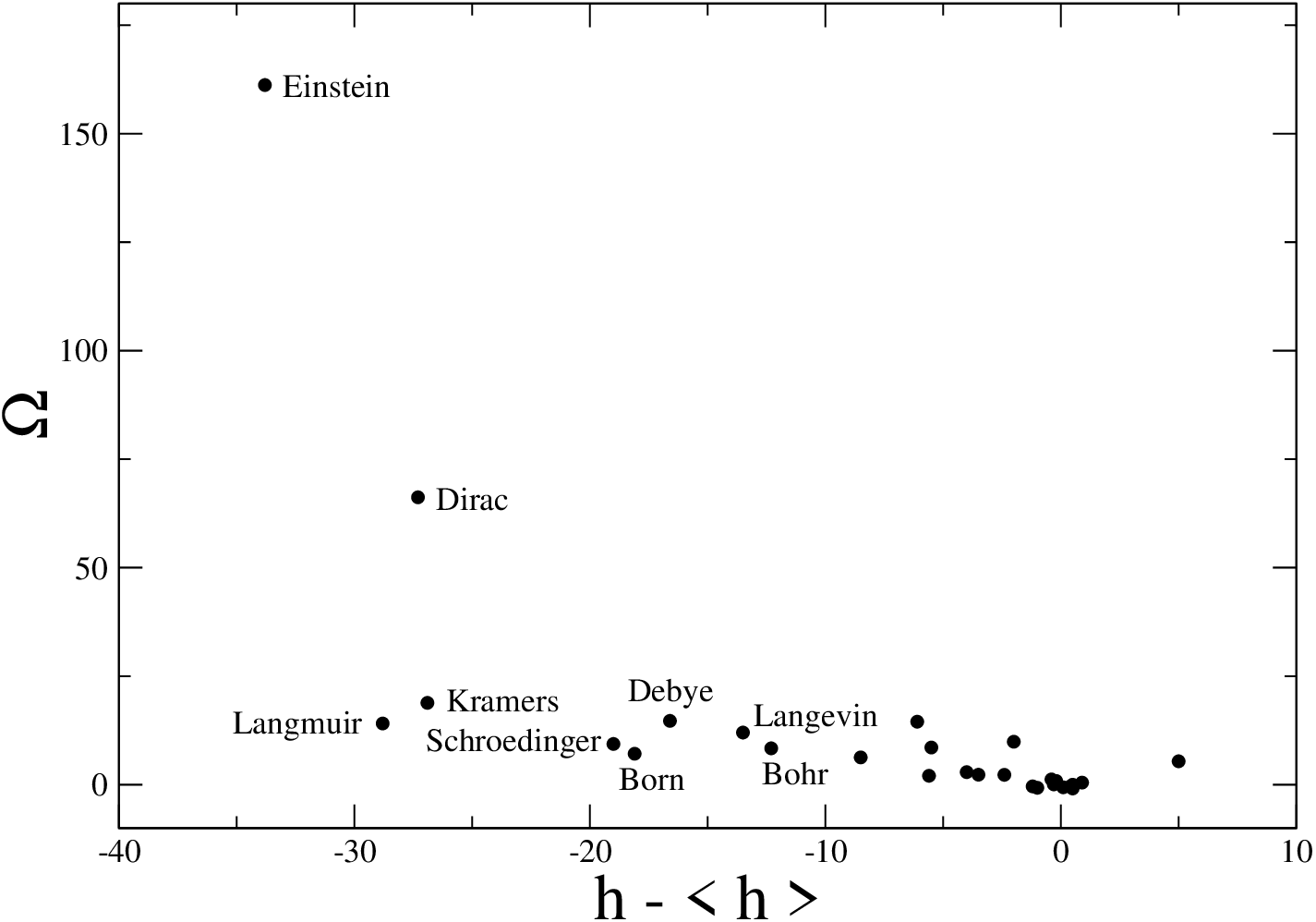}
\caption{Plot of the $\Omega$-index vs. the deviation of the h-index from its average
  value for the participants to the fifth Solvay conference. Some renowned physicists are
  highlighted, with large $\Omega$-index and $h$-index far lower than the average one we should
expect based on the number of citations.}
\label{Fig3}
\end{figure}

To better quantify the difference in the two indicators, we have taken the participants of the
fifth Solvay Conference on Physics in 1927, and compared their $h$ and $\Omega$ indexes, in Table I.
The analysis is then refined in Fig.~\ref{Fig3}, where we plot the $\Omega$-index versus the deviation
of the actual $h$-index from its average value as defined in Eq. (\ref{Eq1}).
It is evident that well-renowed physicists have an $h$-index far smaller than the average one, and $\Omega$
indexes typically larger than unity. The striking figure of the $\Omega$-index for the most
famous physicist of the last century, Albert Einstein, is manifest.

\begin{figure}[t]
\includegraphics[width=1.00\columnwidth, clip=true]{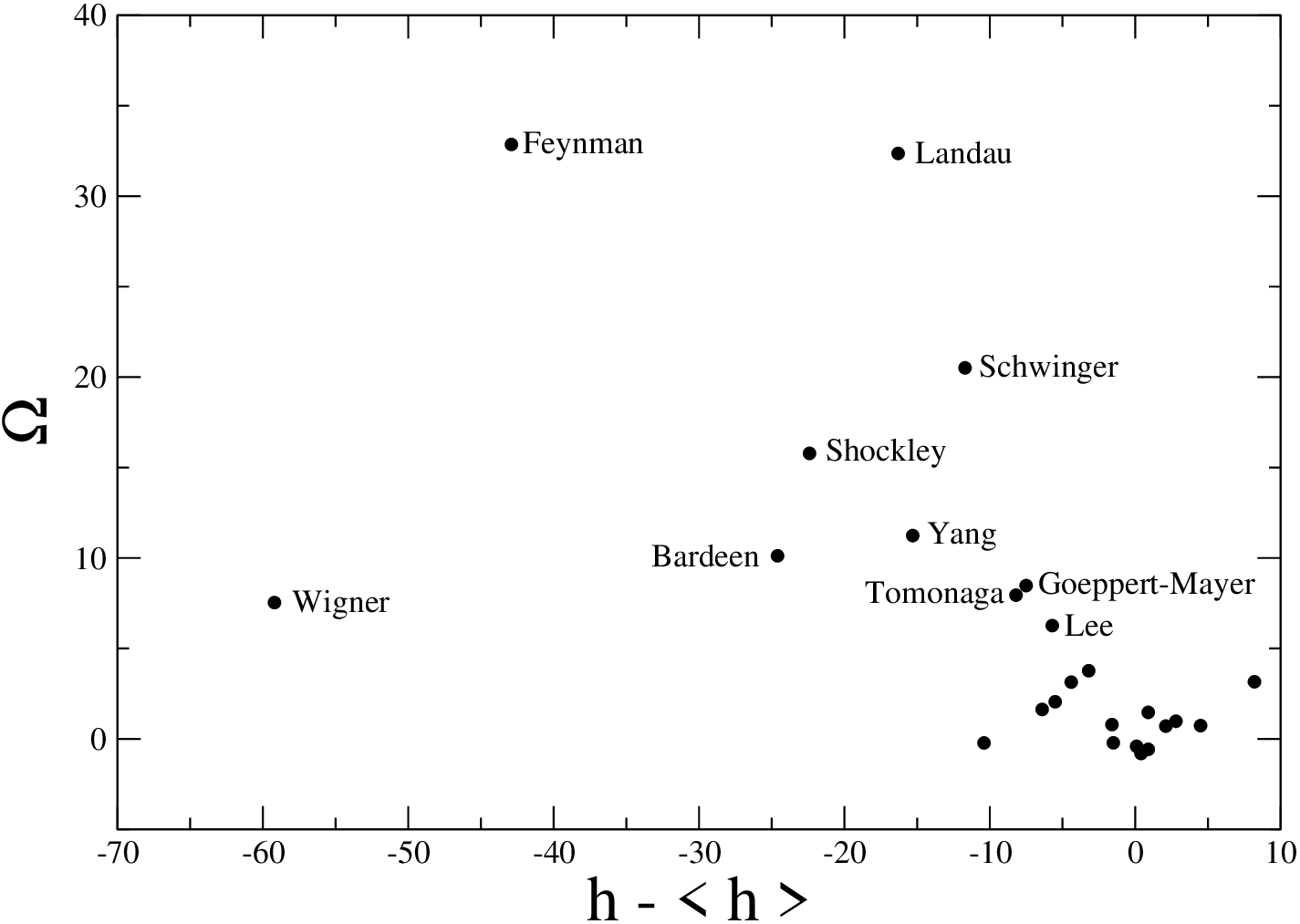}
\caption{Plot of the $\Omega$-index {\it vs.} the deviation of the $h$-index from its average
value for the Nobel laureates in the 1955-1965 timeframe, with highlight of physicists having the
largest $\Omega$-index.}
\label{Fig4}
\end{figure}

The analysis has been also repeated for another relevant period around the mid of the last century.
In Table II we report the relevant quantities for the Nobel laureates in the 1955-1965 period. 
This period has been chosen for a variety of reasons. First, it is one in which there have been
enormous breakthroughs, including new technologies and devices such as lasers, nuclear magnetic
resonance, transistors and personal computers, particle detectors and accelerators, artificial satellites,
fission nuclear power plants and tokamaks, on which we are still working.
Second, on the theory side, this decade starts with the recognition of a puzzle in the high-precision
spectroscopy of hydrogen, the Lamb-shift, and ends with recognition of the theory capable of justifying it,
renormalized quantum electrodynamics; the latter, after fifty years, is still the paradigm against which
to confront any other quantum field theory. An explosion of discoveries and inventions
deserves an analysis in itself, especially in comparison to the relative stagnation of today 
\cite{Jones,Ness,Geman}. It seems evident that a variety of factors, including massive public funding
of basic science, expansion of educational opportunities, and the fallout of the research and development
carried out during WWII, played a major role in this exceptional period. Third, this allows us for testing 
the robustness of the available databases as in the period analyzed in Table I there can be missing
papers, citations and references due to lack of archival data for that early period, in which science
results were also communicated with different channels than the publication in scientific journals.
In Fig.~\ref{Fig4} we repeat the same analysis as in Fig.~\ref{Fig3}, confirming the trend of lower $h$-indexes than
the ones expected on average.
\begin{table}[t]
\begin{center}
{\scriptsize
\begin{tabular}{|l|c|c|c|c|c|l|}
\hline\hline
Author             & $N_{\mathrm{p}}$  & $N_{\mathrm{cit}}$ & $N_{\mathrm{ref}}$ & $h$ & $\langle h \rangle$ & $\Omega$ \\
\hline
Lamb,W.E.         & 123  &  9356   & 2261    & 44  &  48.4   &  3.14         \\
Kusch,P.          &  86  &  3809   &  916    & 36  &  27.8   &  3.16         \\
Bardeen,J.        & 151  & 24696   & 2221    & 54  &  78.6   & 10.12         \\
Brattain,W.H.     &  58  &  2950   &  619    & 24  &  27.2   &  3.77         \\   
Shockley,W.B.     & 127  & 25209   &  1502   & 57  &  79.4   & 15.78         \\
Lee,T.-D.         & 190  & 22920   &  3157   & 70  &  75.7   &  6.26         \\
Yang,C.N.         & 184  & 25803   &  2108   & 65  &  80.3   & 11.24         \\
Cherenkov,P.A.    &  25  &  83     &   428   &  5  &   4.6   & -0.81         \\
Frank,I.          &  54  & 329     &   775   & 10  &   9.1   & -0.58         \\
Tamm,I.E.         &  17  &  95     &   161   &  5  &   4.9   & -0.41         \\
Segr\`e,E.        &  93  & 2752    &  1390   & 29  &  26.2   &  0.98         \\
Chamberlain,O.    & 121  & 2801    &  1616   & 31  &  26.5   &  0.73         \\
Glaser,D.A.       &  66  & 1577    &   924   & 22  &  19.9   &  0.71         \\
Hofstadter,R.     & 223  & 9264    &  3751   & 49  &  48.1   &  1.47         \\
Mossbauer,R.L.    & 92   &  3504   & 1958    & 28  &  29.6   &  0.79         \\
Landau,L.D.       & 74   & 11810   &  354    & 38  &  54.3   & 32.36         \\
Wigner,E.P.       & 151  & 16503   & 1934    & 45  &  64.2   &  7.53         \\
Goeppert-Mayer,M. & 10   &   843   &  89     &  7  &  14.5   &  8.47         \\
Jensen,J.H.D.     & 28   &  1225   &  401    & 12  &  17.5   &  2.06         \\
Basov,N.G.        & 532  &  5637   &  7193   & 36  & 37.5    & -0.22         \\
Prokhorov,A.M.    & 1245 & 10572   & 13608   & 41  & 51.4    & -0.22         \\
Townes,C.H.       & 340  & 19295   & 7345    & 63  & 69.4    &  1.63         \\
Feynman,R.P.      & 77   & 26848   &  793    & 39  & 81.9    & 32.86         \\
Schwinger,J.      & 191  & 28693   & 1334    & 73  & 84.7    & 20.51         \\
Tomonaga,S.-I.    & 31   &  1978   &  221    & 14  & 22.2    &  7.95         \\
\hline
\end{tabular}}
\end{center}
\caption{\label{table2}Rankings of the Nobel laureates in Physics in the 1955-1965 period.}
\end{table}

The $\Omega$-index may even be used to compare with some degree of credibility different
fields of science, or different subfields, without the introduction of rescaling factors \cite{Radicchi}.
Although this needs to be checked with detailed analyses, a plausibility argument in favor
of this point is given by considering two extreme cases based on hypothetical (yet typical)
common-sense figures: the one of a mathematician achieving
60 citations in 10 papers in which there are a total of 20 references, and the case of an experimental 
high-energy physicist who collects 30,000 citations coauthoring a total of 200 papers, each
with 50 references. They will both correspond to an $\Omega=2$, while the average $h$-index will 
be $\langle h \rangle \simeq 3.9$ and $\langle h \rangle \simeq 86.6$, respectively. 
While the evaluation of tenure or promotion of these two scholars within the same committee 
of a school of science seems impossible through the eyes of the $h$-index, and requires the input of 
often highly subjective reference letters from outside reviewers, a rough, first-order comparison is 
possible using the $\Omega$-index. Similar considerations may be applied to a cluster of researchers, for
instance to evaluate the degree of originality of a university department or of a research center. 
Comparison of researchers in different areas of natural and mathematical sciences is crucial, as it
could lead to mitigate compartmentalization and to promote interdisciplinarity. This could lead to
a unified school of sciences, which seems fruitful especially for small and medium size institutions
lacking critical mass of researchers in each individual department.

The $\Omega$-index is also indicator of the vibrancy of a given subfield of science.
Summing over all the references in a subfield at a given year $y$, $\Omega_y$, and comparing 
this to the total number of citations of the following year $\Omega_{y+1}$, one may have a figure of the rate 
of growth of the subfield, with a subfield contracting ($\Omega_{y+1}<\Omega_{y}$), in a steady state 
($\Omega_{y+1}=\Omega_y$), or expanding ($\Omega_{y+1}>\Omega_y$). 

While the $h$-index is monotonic in time, the $\Omega$-index is not, and prolific production of 
incremental papers following a breakthrough, or of secondary scholarship like review papers, 
will tend to decrease the $\Omega$-index. At the same time, the $h$-index is bounded to
the total number of published papers after a researcher ceases the scientific activity, while the
$\Omega$-index can grow indefinitely. In an apparent paradoxical fashion, as an alternative 
to produce other breakthrough papers, the best strategy to maintain or increase a large 
$\Omega$-index is to stop publishing for a while, just focusing on the problem at hand or
studying to tackle a completely new problem. This seems to be what common sense suggest to
secure a series of breakthrough, high-risk/high-reward, results. 

The main drawback of the $\Omega$-index seems to be the risk that some authors will tend to
minimize as much as possible references to previous work, which will require a more focused 
action from the side of referees and editors. On the other hand, the expected dramatic reduction
in the number of published papers, in the case the $\Omega$-index will be considered, will 
allow both authors and referees to read more carefully the existing literature.
It is then not unlikely to imagine that a submitted paper will be carefully scrutinized by a
set of four-five reviewers for instance, focusing more attention on the careful, fair citation of
former contributions. Moreover, the $\Omega$-index seems to penalize junior researchers, as typically
it takes several years before reaching at least the break-even point corresponding to $\Omega=0$ (with
everybody starting from an initial $\Omega=-1$). It should be remarked, though, that similar considerations
apply to the $h$-index, which has been suggested to be used for relatively senior researchers. Overall, in
spite of sophisticated indexes, the best indicator at the junior level is to provide a
self-selected number of few (say a maximum of three) papers, on which any knowledgeable
committee can focus the attention to the greatest possible depth.

\section{Conclusions}

We have discussed two relevant drawbacks of the $h$-index if used to identify originality in scientific work.
First, in an average sense its value is merely determined once the number of citations is known.
Second, deviations from such an average value indicate that lower values of the $h$-index correspond 
to a small set of highly cited papers, while higher values are indicative of cumulative, medium-impact science. 
We have then introduced a simple indicator, the $\Omega$-index, which appears to better approximate the
perceived feeling of a contribution being more or less original. This is corroborated by the analysis of
data available on two historical periods in which highly creative work in Physics has been performed, one
corresponding to the birth of the ``new'' quantum mechanics, the other to the enormous number of breakthroughs
experienced in the Fifties. The analysis of the two samples shows that, by adopting the $h$-index as the
unique indicator for hiring and promotion, as could happen these days, most of the prominent scientist
in the lists would have {\it not} been considered for academic positions even with {\it post-mortem} applications. 
The dynamics of optimization of the $\Omega$-index involves more feedback and nonlinearity, and does
not end up encouraging unlimited proliferation of papers and review articles.

Finally, we would like to briefly comment on how the evolution of science may be affected by 
the active use of different indicators. A community aiming to maximize the $h$-index will 
try to be as prolific as possible in terms of papers and review articles, favoring authorship 
fusion processes, large collaborations, and discouraging changes of subfields. 
Demonstrational physics, in which a well-established theory is repeatedly confirmed in the laboratory for 
technological purposes, and applied sciences at large, may tend to benefit more from the $h$-index. 
Conversely, a community willing to maximize the $\Omega$-index will carefully consider the need 
to publish new papers, and to cite former papers, will favor the opening of uncharted territories 
for which former references may be scarce or barely existing, will discourage review papers 
unless they fill a clearly perceived gap, and will mitigate the incentive for authorship fusion. 
More theoretically-oriented work, especially in basic mathematics and surrounding fields, and 
pioneering experimental work in fields for which the theoretical counterpart is not settled or
not even existing, could greatly benefit from the adoption of the $\Omega$-index. 
At a strictly individual level, leadership is more recognized with the $h$-index, while pioneering capabilities 
are more recognized with the $\Omega$-index. This suggests that the $\Omega$-index 
should be considered as an alternative, or at least complementary, to the $h$-index, in 
all subfields of basic/long-term return science, since a blind, pedantic application of the $h$-index 
alone could result in devastating effects on the entire scientific ecosystem. 

\acknowledgments
I am grateful to Felix Beaudoin, Leigh M. Norris, Alexander R. H. Smith, Bala Sundaram and Lorenza Viola
for a critical reading of the manuscript.

\end{document}